\documentclass[%
 aps,prl,letterpaper,floatfix,
 preprintnumbers,
 reprint,
 amsmath,amssymb,
 showpacs,showkeys,
 superscriptaddress,
]{revtex4-1}

\usepackage{graphicx}
\usepackage{bm}
\usepackage[dvipdfm]{hyperref}
\hypersetup{bookmarksnumbered, pdfpagemode=UseOutlines, 
pdfauthor={I.\ J.\ Vera-Marun}, 
pdftitle={Spin heat accumulation induced by tunneling from a ferromagnet}, pdfdisplaydoctitle, 
colorlinks=true, citecolor=blue, filecolor=blue, linkcolor=blue, urlcolor=blue}

\begin{document}

\title{Spin heat accumulation induced by tunneling from a ferromagnet}
\author{I.\ J.\ \surname{Vera-Marun}}
	\email[e-mail: ]{i.j.vera.marun@rug.nl} 
\author{B.\ J.\ \surname{van Wees}}
\affiliation{Physics of Nanodevices, Zernike Institute for Advanced Materials, University of Groningen, Nijenborgh 4, 9747 AG Groningen, The Netherlands}
\author{R.\ Jansen}
\affiliation{National Institute of Advanced Industrial Science and Technology (AIST), Spintronics Research Center, Tsukuba, Ibaraki, 305-8568, Japan}

\date{\today}

\begin{abstract}
An electric current from a ferromagnet into a non-magnetic material can induce a spin-dependent electron temperature. Here it is shown that this spin heat accumulation, when created by tunneling from a ferromagnet, produces a non-negligible voltage signal that is comparable to that due to the coexisting electrical spin accumulation and can give a different Hanle spin precession signature. The effect is governed by the spin polarization of the Peltier coefficient of the tunnel contact, its Seebeck coefficient, and the spin heat resistance of the non-magnetic material, which is related to the electrical spin resistance by a spin-Wiedemann-Franz law. Moreover, spin heat injection is subject to a heat conductivity mismatch that is overcome if the tunnel interface has a sufficiently large resistance.
\end{abstract}


\maketitle

Creation and detection of spin information are at the heart of spintronics, the study and use of spin degrees of freedom \cite{zutic_spintronics:_2004}. Electronic spin transport is described by a two-channel model where transport is separately considered for each spin ($\sigma = \uparrow , \downarrow$) population \cite{mott_electrical_1936, fert_two-current_1968, van_son_boundary_1987}. When a current is applied between a ferromagnetic contact and a non-magnetic material, it induces a spin accumulation $\Delta\mu = \mu^\uparrow - \mu^\downarrow$ described by a splitting of the electrochemical potentials $\mu^{\sigma}$ of the two spin channels in the non-magnetic material \cite{valet_theory_1993, jedema_electrical_2002, lou_electrical_2006, dash_electrical_2009}. Direct electrical detection of the spin accumulation is achieved via the Hanle effect, where a magnetic field induces spin precession and suppresses $\Delta\mu$, giving a measurable voltage signal. Interestingly, spin current in ferromagnetic tunnel contacts can be created both by an electrical bias \cite{tedrow_spin-dependent_1971, meservey_spin-polarized_1994} or by a thermal bias \cite{le_breton_thermal_2011, jansen_thermal_2012}. The latter approach of thermal spin injection is possible due to the spin dependence of thermoelectric properties in magnetic materials and nanodevices, which lead to interactions between spin and heat transport currently studied in the field of spin caloritronics \cite{hatami_thermoelectric_2009, bauer_spin_2012}. This raises the question: do Hanle measurements only detect a difference in electrochemical potentials $\mu^{\sigma}$, as hitherto assumed, or also a difference in temperatures $T^{\sigma}$ between the two spin channels?

In this work we address the creation and detection of a spin heat accumulation $\Delta T_{s} = T^{\uparrow}-T^{\downarrow}$ in a non-magnetic material via a ferromagnetic tunnel contact. It is considered here that tunneling transport is accompanied by a spin-dependent heat flow if the Peltier coefficient of the tunnel contact depends on spin, and that this produces a spin heat accumulation and an additional contribution to the voltage signal in a Hanle measurement. Spin heat accumulation is a concept previously studied theoretically within the context of metallic spin-valve structures \cite{giazotto_nonequilibrium_2007, hatami_thermal_2007, heikkila_spin_2010} and only very recently it has been observed as a spin-dependent heat conductance in metallic current-perpendicular-to-plane spin-valve nanopillars \cite{dejene_spin_2013_published}. Here, we provide an explicit evaluation for the spin heat accumulation at the tunnel interface between a ferromagnet and a non-magnetic material. Notably, we introduce the notion of an associated heat conductivity mismatch, similar to that for spin accumulation \cite{schmidt_fundamental_2000, rashba_theory_2000, fert_conditions_2001, takahashi_spin_2003, maassen_contact-induced_2012}, which limits the magnitude of spin heat accumulation and can be overcome with the tunnel interface. Most importantly, we show that the widely employed Hanle measurement to detect spin accumulation has another contribution from the spin heat accumulation that can be comparable in magnitude and has a line width set by the spin heat relaxation time. It cannot be neglected a priori and needs to be considered for a correct interpretation of experimental data.

\begin{figure}[tbp]
\includegraphics*[
width=0.9\columnwidth, 
trim = 0mm 1mm 0mm 0mm, clip]{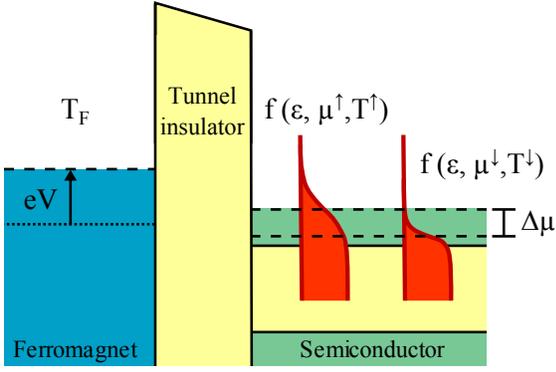}
\caption{\label{fig:one}
Energy band diagram of a ferromagnet-insulator-semiconductor tunnel junction. The electrons in the semiconductor have spin-dependent temperatures $T^{\uparrow}$ and $T^{\downarrow}$, whereas those in the ferromagnet are at $T_{F}$. Also, a spin accumulation exists in the semiconductor, described by a spin splitting $\Delta\mu = \mu^{\uparrow}-\mu^{\downarrow}$ of the electrochemical potential. The distribution functions are indicated by the red lines.}
\end{figure}

We consider the case of a three-terminal geometry, where the same contact is used for driving an electrical current and measuring the voltage signal, as commonly used for spin injection into semiconductors \cite{lou_electrical_2006, dash_electrical_2009}, although the basic physics also applies to other device geometries, such as the non-local one. Such a tunnel junction with a ferromagnetic electrode and a non-magnetic semiconductor electrode is depicted in Fig.~\ref{fig:one}. We describe each spin population in the non-magnetic material by a Fermi-Dirac distribution with spin-dependent temperatures $T^{\uparrow}$ and $T^{\downarrow}$. This is strictly valid only when thermalization within each spin channel is sufficiently fast compared to energy exchange between the spin channels. In general, the distributions could be non-thermal and we should regard $T^\sigma$ as effective temperatures \cite{heikkila_spin_2010, dejene_spin_2013_published}. For the ferromagnet we assume negligible spin and spin heat accumulations due to stronger spin-flip and inelastic scattering processes, so both spin channels are equilibrated at $T_F$. We define an average electron temperature $T_0 = (T^{\uparrow} + T^{\downarrow})/2$ in the non-magnetic material, and a temperature difference $\Delta T_0 = T_0 - T_F$ across the contact. The charge tunnel currents $I^{\sigma}$ and the \emph{electronic} heat currents $I^{Q,\sigma}$ for each spin channel are then given by \cite{[{Eqs. (1) and (2) are similar to previous work, see }][{, except for the newly added terms containing the spin heat accumulation $\Delta T_{s}$.}] jansen_thermal_2012} 
\begin{eqnarray}
I^{\sigma} &=& G^{\sigma}\left(V\mp\frac{\Delta\mu}{2\,e}\right) + L^{\sigma} \left(\Delta T_0\pm\frac{\Delta T_{s}}{2}\right) \label{eq1}\\
I^{Q,\sigma} &=& -\kappa^{el,\sigma} \! \left( \Delta T_0\pm\frac{\Delta T_{s}}{2} \right) + G^{\sigma}S^{\sigma}T_0 \! \left( V\mp\frac{\Delta\mu}{2\,e}\right) \label{eq3}
\end{eqnarray}
with $V$ the voltage across the junction and $\kappa^{el,\sigma}$ the \emph{electronic} heat conductance of the tunnel barrier in units of [Wm$^{-2}$K$^{-1}$]. Charge currents $I^\sigma$ are in units of [Am$^{-2}$], conductances $G^\sigma$ in [$\Omega^{-1}\text{m}^{-2}$] and heat currents $I^{Q,\sigma}$ in [Wm$^{-2}$]. By definition, $I>0$ and $I^Q>0$ correspond to electron flow and heat flow, respectively, from the ferromagnet to the semiconductor. Furthermore, the Onsager coefficient $L$ (thermoelectric conductance) is positive when the conductance below the Fermi energy is larger than that above it, so the spin-dependent Seebeck coefficient $S^\sigma=-L^\sigma/G^\sigma<0$ for hole-like transport \footnote{This convention is consistent with previous work on Seebeck spin tunneling \cite{le_breton_thermal_2011, jansen_thermal_2012}, but opposite to that commonly used in the field of thermoelectrics.}.

First we proceed to find the spin current and the spin heat current injected into the non-magnetic material. The spin current  $I_s=I^{\uparrow}-I^{\downarrow}$, and the charge current $I=I^{\uparrow}+I^{\downarrow}$, are obtained from Eq.~(\ref{eq1})
\begin{eqnarray}
I &=& G\,V - P_G\,G\,\left(\frac{\Delta\mu}{2\,e}\right) + L \,\Delta T_0 + P_L\,L\,\left(\frac{\Delta T_{s}}{2}\right) \label{eq5} \\
I_s &=& P_G\,G\,V - G \left(\frac{\Delta\mu}{2\,e}\right) + P_L\,L\,\Delta T_0 + L\,\left(\frac{\Delta T_{s}}{2}\right) \label{eq6}
\end{eqnarray}
where we have defined the total conductances $G = G^{\uparrow}+G^{\downarrow}$ and $L = L^{\uparrow}+L^{\downarrow}$, and their spin polarizations $P_G = (G^{\uparrow}-G^{\downarrow})/(G^{\uparrow}+G^{\downarrow})$ and $P_L = (L^{\uparrow}-L^{\downarrow})/(L^{\uparrow}+L^{\downarrow})$.

The \emph{total} heat current $I^Q=I^{Q,\uparrow} + I^{Q,\downarrow} + I^{Q,ph}$ contains, in addition to the heat flow by electrons, a dominant contribution due to phonon transport across the tunnel contact. It is given by $I^{Q,ph} = -\kappa^{ph}\,(T^{ph}-T_{F})$, where $\kappa^{ph}$ is the phonon heat conductance of the barrier (usually dominated by the interfaces \cite{ju_nanoscale_2005}), $T^{ph}$ is the phonon temperature in the non-magnetic electrode, and we assume that in the ferromagnet phonons and electrons are fully equilibrated at $T_{F}$. Phonon heat flow is not parameterized by the spin variable and does not contribute to the spin heat current across the barrier. Thus, the injected spin heat current $I_s^Q = I^{Q,\uparrow} - I^{Q,\downarrow}$ is only due to the electrons and can be obtained from Eq.~(\ref{eq3})
\begin{eqnarray}
I_s^Q &=& -P_{\kappa}^{el}\,\kappa^{el}\,\Delta T_0 - \kappa^{el}\,\left(\frac{\Delta T_{s}}{2}\right) \nonumber\\
 &&- P_L\,L\,T_0\,V + L\,T_0\,\left(\frac{\Delta\mu}{2\,e}\right) \label{eq8}
\end{eqnarray}
where we have defined the total electronic heat conductance of the tunnel contact $\kappa^{el} = \kappa^{el,\uparrow} + \kappa^{el,\downarrow}$, its polarization $P_{\kappa}^{el} = (\kappa^{el,\uparrow} - \kappa^{el,\downarrow})/(\kappa^{el,\uparrow} + \kappa^{el,\downarrow})$ and used the relation $S^{\sigma}G^{\sigma}=-L^{\sigma}$.

\begin{figure}[tbp]
\includegraphics*[trim = 0mm 0mm 0mm 0mm, clip]{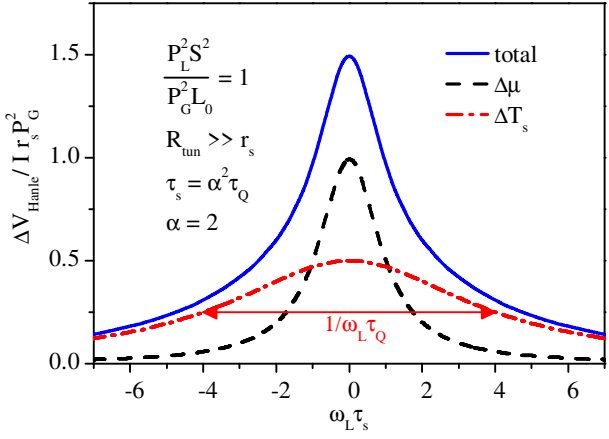}
\caption{\label{fig:two}
Hanle signal in a three-terminal configuration and the contributions corresponding to spin accumulation $\Delta \mu$ and spin heat accumulation $\Delta T_{s}$. The relative magnitude of the signals is obtained from Eq.~(\ref{eq23}), assuming $(P_{L} S)^2/(P_G^2 L_0) = 1$, $\alpha = 2$, and $R_{tun} \gg r_s$. We consider a spin heat relaxation time $\tau_Q < \tau_s$,  given by $\tau_s / \tau_Q = (\lambda_s / \lambda_Q)^2 = \alpha^2$.}
\end{figure}

Now we can evaluate the contribution of the spin heat accumulation to a Hanle measurement.
Typically, a Hanle measurement involves the application of a constant electrical current $I$ at the tunnel junction while spin precession in a magnetic field $B$ perpendicular to the injected spins causes $\Delta\mu$ to go to zero. The decrease in $\Delta\mu$ depends on the product of the spin-relaxation time $\tau_{s}$ and the Larmor frequency $\omega_{L} = g \mu_B B / \hbar$, with $g$ the Land\'e g-factor, $\mu_B$ the Bohr magneton, and $\hbar$ the reduced Planck constant. Importantly, spin precession would also cause $\Delta T_{s}$ to go to zero. This can be understood by considering a packet of hot spins polarized along $+x$ and an equal amount of cold spins polarized along $-x$. A perpendicular field along $z$ causes a precession of each spin in the $x$--$y$ plane and thereby a periodic oscillation of the temperature of the electrons with spin pointing along $+x$ (or $-x$) \cite{note_1}. If the precession frequency $\omega_{L}$ is much larger that the inverse of the time constant $\tau_Q$, associated with relaxation of the spin heat accumulation, then the time-average of $\Delta T_{s}$ goes to zero. Therefore, for such an electrically driven junction, and assuming that the spin-averaged temperatures of the electrodes remain constant, the corresponding Hanle signal $\Delta V_{Hanle} = V - V|_{\Delta\mu, \Delta T_{s} \rightarrow 0}$ can be obtained from Eq.~(\ref{eq5})
\begin{equation}
\Delta V_{Hanle} = \left(\frac{P_G}{2\,e}\right)\Delta\mu + \left(\frac{P_L\,S}{2}\right)\Delta T_{s} \label{eq10}
\end{equation}
In addition to the well-known Hanle signal arising from the spin accumulation $\Delta \mu$, there is a second up to now neglected contribution due to the spin heat accumulation $\Delta T_{s}$. Note that the Hanle curve $\Delta V_{Hanle}$~\text{vs} ~$B$, due to suppression of $\Delta \mu$, is Lorentzian \cite{zutic_spintronics:_2004} and has a width that is inversely proportional to the spin-relaxation time $\tau_{s}$. Similarly, we expect that the suppression of $\Delta T_{s}$ yields a Lorentzian Hanle curve having a width that is inversely proportional to the spin heat relaxation time $\tau_{Q}$, such that $\Delta T_{s}(B) \propto ( \, 1+ ( \omega_{L} \tau_{Q} )^2 \, )^{-1}$. If $\tau_Q$ is sufficiently different than $\tau_{s}$, then the total Hanle signal will consist of two superimposed Hanle curves with different widths, as depicted in Fig.~\ref{fig:two}. We remark that the latter directly follows from the sum of two independent contributions to the voltage, conform Eq.~(\ref{eq10}). Interestingly, if a spin heat accumulation is present, interpreting the total Hanle signal purely in terms of a spin accumulation would lead to an underestimation of the spin-relaxation time $\tau_s$.

Next, we evaluate the created spin accumulation $\Delta\mu$ and spin heat accumulation $\Delta T_s$ in the non-magnetic material. We consider a steady-state condition in which the spin current $I_s$ injected by tunneling is balanced by the spin current due to spin relaxation processes in the material \cite{valet_theory_1993}, occurring over the spatial extent of $\Delta\mu$. Similarly, the spin heat current $I_s^Q$ injected by tunneling is balanced by the heat current between the two spin populations due to spin relaxation \emph{and} inelastic scattering processes \cite{heikkila_spin_2010}, occurring over the spatial extent of $\Delta T_s$. To relate accumulations and  injected currents we define a spin resistance $r_s$ and a spin heat resistance $r_s^Q$ of the non-magnetic material
\begin{eqnarray}
\Delta\mu &=& 2\,e\,I_s\,r_s \label{eq11} \\
\Delta T_{s} &=& 2\,I_s^Q\,r_s^Q \label{eq12}
\end{eqnarray}
where $r_s$ is a phenomenological parameter that describes the conversion of the spin current $I_s$ injected by tunneling, into a spin accumulation with a value of $\Delta\mu$ right at the tunnel interface, as before \cite{jansen_silicon_2012}. This definition does not require us to assume any  specific profile of the spin accumulation in the non-magnetic material. If we do assume that the spin accumulation decays exponentially away from the tunnel interface with a spin-relaxation length $\lambda_s$ \cite{valet_theory_1993}, then the spin resistance per unit area is $r_s = \rho \, \lambda_s$, where $\rho$ is the resistivity of the non-magnetic material \cite{fert_conditions_2001, fert_semiconductors_2007, jaffres_spin_2010}. Similarly, the parameter $r_s^Q$ is defined in terms of the injected spin heat current $I_s^Q$ and the spin heat accumulation $\Delta T_{s}$ right at the tunnel interface.

Using the definition above for spin heat resistance we can obtain $\Delta T_s$ from Eqs.~(\ref{eq8}) and (\ref{eq12})
\begin{eqnarray}
\Delta T_s &=& \left\{\frac{r_s^Q\,R_{tun}^{Q,el}}{R_{tun}^{Q,el} + r_s^Q}\right\} \nonumber\\*
&&\times\left[\frac{(2\,P_L\,V - \Delta\mu/e)}{R_{tun}}\,S\,T_0 -\frac{2\,P_{\kappa}^{el}}{R_{tun}^{Q,el}}\,\Delta T_0 \,\right] \label{eq14}
\end{eqnarray}
where $R_{tun}=1/G$ is the tunnel resistance, and $R_{tun}^{Q,el} = 1/\kappa^{el}$ is the \emph{electronic} thermal resistance of the tunnel barrier. The two terms within the square brackets represent the sources of spin-dependent heat flow. The first one is due to the spin heat current that accompanies the charge current across the tunnel contact, governed by the spin polarization $P_L$ of the Peltier coefficient \cite{flipse_direct_2012}. The second term is present when there is a temperature bias across the junction, driving a spin-dependent heat flow if the heat conductance of the tunnel barrier is spin dependent ($P_{\kappa}^{el}\neq0$) \cite{dejene_spin_2013_published}. If transport through the tunnel barrier is elastic, the tunnel resistance and the electronic thermal resistance of the tunnel barrier are interrelated by the Wiedemann-Franz law \cite{stovneng_thermopower_1990}
\begin{equation}
R_{tun}^{Q,el} = \frac{1}{\kappa^{el}} = \frac{1}{L_0\,T_0\,G} = \frac{R_{tun}}{L_0\,T_0} \label{eq18}
\end{equation}
with $L_0 \approx 2.45\times10^{-8}\,V^2\,K^{-2}$ the Lorentz number. This allows us to estimate the magnitude of $R_{tun}^{Q,el}$ and, together with Eqs.~(\ref{eq5}) and (\ref{eq14}), to obtain an explicit evaluation of the spin heat accumulation $\Delta T_s$ in terms of the driving current $I$. Furthermore, we can also obtain an explicit evaluation for the spin accumulation $\Delta\mu$ from Eqs.~(\ref{eq5}), (\ref{eq6}) and (\ref{eq11}). The resulting expressions are
\begin{widetext}
\begin{eqnarray}
\Delta\mu &=& \left\{\frac{r_s\,R_{tun}}{R_{tun} + (1 - P_G^2)r_s}\right\} \left( \frac{e}{R_{tun}} \right) \left[\,2\,P_G\,R_{tun}\,I + (1 - P_G^2)\,(S^{\uparrow} - S^{\downarrow})\,\Delta T_0 - (1-P_L\,P_G)\,S\,\Delta T_s \, \right] \label{eq13}\\
\Delta T_s &=& \left\{\frac{r_s^Q\,R_{tun}^{Q,el}}{R_{tun}^{Q,el} + \left(1-\frac{P_L^2\,S^2}{L_0}\right)r_s^Q}\right\} \left(\frac{S\,T_0}{R_{tun}}\right) \left[2\,P_L\,R_{tun}I - (1 - P_L\,P_G)\,\frac{\Delta\mu}{e} + 2\left(P_L - P_{\kappa}^{el}\frac{L_0}{S^2}\right)S\,\Delta T_0 \, \right] \label{eq20}
\end{eqnarray}
\end{widetext}

Note the similarity among the terms between curly brackets in Eqs.~(\ref{eq13}) and (\ref{eq20}). The term in Eq.~(\ref{eq13}) corresponds to the known issue of conductivity mismatch \cite{schmidt_fundamental_2000, rashba_theory_2000, fert_conditions_2001, takahashi_spin_2003, maassen_contact-induced_2012} which limits the magnitude of the spin accumulation due to back flow of the spins into the ferromagnet when $R_{tun} < r_s$. Remarkably, the term in Eq.~(\ref{eq20}) alludes to an analogous notion of a heat conductivity mismatch: if the heat resistance $R_{tun}^{Q,el}$ of the tunnel barrier is smaller than the spin heat resistance $r_s^Q$ of the non-magnetic material, then the spin heat accumulation is reduced by back flow of the spin heat into the ferromagnet. In order to overcome the heat conductivity mismatch, one needs to fulfill the condition $R_{tun}^{Q,el} \gg r_s^Q$. Note that this is governed by the electronic heat resistance of the tunnel contact, since phonons cannot transport spin heat across the tunnel barrier. This concept is crucial to the creation of a large $\Delta T_s$, which in recent experimental work was limited by highly transparent metallic contacts \cite{dejene_spin_2013_published}. 

Finally, we evaluate the magnitude of the spin heat accumulation and its corresponding contribution to the Hanle signal. Both are fully described by Eqs.~(\ref{eq10}), (\ref{eq13}) and (\ref{eq20}). In the following, we avoid the in general lengthy solutions and proceed to make a few practical simplifications. First, we consider a junction that is driven \emph{electrically}, not thermally, so that the spin accumulation is dominated by electrical spin injection. We therefore neglect the second term (due to Seebeck spin tunneling \cite{le_breton_thermal_2011}) and third term (due to a non-zero spin heat accumulation $\Delta T_s$) in Eq.~(\ref{eq13}), and retain only the first term proportional to $I$. Note that the last term $S\,\Delta T_s$ is expected to be smaller than 1~mV, and thus small compared to $R_{tun}\,I$. Similarly, the dominant term in Eq.~(\ref{eq20}) for $\Delta T_s$ is the first term, proportional to $I$ (due to the spin-dependent Peltier effect). This is valid since $\Delta\mu / e \ll R_{tun}\,I$,
and $\Delta T_0$ is typically smaller than $S\,R_{tun}\,I/L_0 \approx 40\,K$ for reasonable values of $S=100\,\mu VK^{-1}$ and $R_{tun}\,I=100\,mV$.

Still, there is one parameter with an unknown magnitude, the spin heat resistance of the non-magnetic material $r_s^Q$. It can be expressed as \cite{note_1} 
\begin{equation}
r_s^Q = \frac{r_s}{L_0\,T_0} \times \frac{1}{\alpha} \label{eq22}
\end{equation}
This constitutes a type of spin-Wiedemann-Franz law, relating the electronic spin resistance to the spin heat resistance.
The parameter $\alpha$ takes into account the inelastic scattering processes occurring within the non-magnetic material which increase the interspin heat exchange. These microscopic processes, described in Ref.~\onlinecite{heikkila_spin_2010}, correspond to electron-electron and electron-phonon interactions that cause relaxation of the spin heat accumulation and decrease $\tau_Q$. We remark that $\alpha$ is related to the concept of a spin heat relaxation length $\lambda_Q = \sqrt{D\,\tau_Q}$ in the regime of diffusive (heat) transport, with the corresponding diffusion constant $D$ being the same as that for charge transport since \emph{electronic} heat transport is associated with electrons only.
In the case of an exponentially decaying spin (heat) accumulation it follows that $\alpha = \lambda_s / \lambda_Q$, which in recent work \cite{dejene_spin_2013_published} has been estimated to be $\lambda_s / \lambda_Q \approx 5$ for Cu at room temperature. Therefore, $\alpha$ can indeed be of order one. In general we may expect $\alpha > 1$, although the definition \cite{note_1} of $\alpha$ does not preclude a value smaller than unity. At low temperatures inelastic scattering processes are reduced \cite{heikkila_spin_2010}, so we expect elastic spin-flip scattering to be the dominant spin relaxation mechanism and $\alpha \rightarrow 1$. Using this result, we finally obtain for the Hanle signal
\begin{eqnarray}
\frac{\Delta V_{Hanle}}{I} = r_s\left(P_G\right)^2\left\{\frac{R_{tun}}{R_{tun} + (1 - P_G^2)\,r_s}\right\} \nonumber \\*
+ \frac{r_s}{\alpha}\left(\frac{P_L^2\,S^2}{L_0}\right)\left\{\frac{R_{tun}}{R_{tun} + \left(1-\frac{P_L^2\,S^2}{L_0}\right)\frac{r_s}{\alpha}}\right\} \label{eq23}
\end{eqnarray}
where the first term is due to $\Delta\mu$ and the second term is due to $\Delta T_s$.

The relative magnitude of the two contributions to the Hanle signal is governed by the ratio $P_L/P_G$, by $\alpha$, and by the factor $S^2/L_0$. The latter is unity for $S\simeq 157\,\mu$V/K, which, in the Sommerfeld approximation, is the maximum value with a non-negative entropy production, as required by the second law of thermodynamics \cite{guttman_thermoelectric_1995}. Previous work has shown that the Seebeck coefficient of a tunnel junction is indeed in the order of $100\,\mu$V/K \cite{leavens_vacuum_1987, marschall_charge_1993, le_breton_thermal_2011, walter_seebeck_2011}, and for the case of ferromagnetic electrodes is enhanced by magnons \cite{mccann_giant_2002, mccann_tunnel_2004}. If $P_L \sim P_G$, the Hanle signals due to $\Delta T_{s}$ and $\Delta\mu$ are then comparable in magnitude. Interestingly, both contributions always show the same sign, because there are only quadratic terms.

We conclude that the spin heat accumulation can make a significant contribution to the Hanle signal that cannot be neglected a priori. In principle, it can even be larger than the regular Hanle signal from the spin accumulation if $P_L > P_G$, which can occur when $S^{\uparrow} \neq S^{\downarrow}$ since we have $P_L = P_G - (1-P_G^2)(S^{\uparrow}-S^{\downarrow})/(2\,S)$. However, the resulting enhancement of the Hanle signal is not sufficiently large to explain recent experiments in which Hanle signals that scale with the tunnel resistance and are many orders of magnitude larger than predicted by theory are observed
\cite{tran_enhancement_2009, 
li_electrical_2011, jeon_electrical_2011, uemura_critical_2012,
sharma_anomalous_2014, txoperena_how_2013, han_spin_2013},
as recently reviewed \cite{jansen_silicon_2012-nmat, jansen_silicon_2012}. It would require $\alpha \ll 1$ by several orders of magnitude. To estimate the magnitude of the spin heat accumulation we use Eq.~(\ref{eq20}) in the regime without heat conductivity mismatch ($R_{tun}^Q \gg r_s^Q$), and retain only the leading term, so we obtain
$\Delta T_s = 
2 \,r_s \,S \,P_L \,I / L_0 \,\alpha $. 
For reasonable parameters at room temperature of $r_s=25\,\Omega \mu m^2$, $S=100\,\mu V/K$, and $P_L=50\%$, a modest current density of $10^{7}\,A m^{-2}$ would yield a spin heat accumulation of $\Delta T_s \approx 1\,K$ (for $\alpha=1$).

The creation, manipulation, and detection of spin heat accumulation is a subject that is still in its infancy. The realization that it contributes to Hanle measurements, given the spin dependence of Peltier coefficients in magnetic tunnel contacts, makes it a non-negligible factor that needs to be taken into account in current studies of spin injection. It affects the magnitude and width of the Hanle curve and its variation with temperature, and is also expected to be present in (non)local spin-valve measurements.
The analogy between spin and heat transport, here made explicit by the concept of a heat conductivity mismatch, opens opportunities to address fundamental questions about the relaxation of a spin heat accumulation. How do inelastic processes affect the magnitude of the parameter $\alpha$ in the spin-Wiedemann-Franz law and the width of the Hanle curve, and how is this behavior modified under the presence of a strong spin-orbit coupling. And is it possible that $\alpha < 1$, meaning that the spin heat relaxes slower than the spin accumulation?

\begin{acknowledgments}
IJVM thanks F.~K.~Dejene for useful discussions. This work was financially supported by the Zernike Institute for Advanced Materials and the Netherlands Organisation for Scientific Research (NWO).
\end{acknowledgments}

\bibliography{Zotero_v11}

\newpage


\section{Supplemental Material}

\section{Spin-Wiedemann-Franz law}

Here we develop in more detail the concept of a spin-Wiedemann-Franz law used in the main text. First, we prove mathematically the aforementioned law for the case of a spin relaxation dominated by elastic spin-flip scattering. Next, we include inelastic relaxation processes and define the parameter $\alpha$. Finally, we show that for elastic spin-flip scattering a non-zero spin heat accumulation does not induce nor change the spin accumulation.

\subsection{Spin heat resistance and Wiedemann-Franz law}

The standard way to compute the steady state value of the spin accumulation $\Delta\mu$ is to balance the net amount of spins injected per unit time $t$, with the loss of spins due to spin relaxation in the non-magnetic material. The latter is governed by the spin resistance $r_s$, for which $r_s = \rho \, \lambda_s$ is obtained if we assume that $\Delta\mu$ decays exponentially away from the tunnel interface with a spin-relaxation length $\lambda_s$ \cite{fert_conditions_2001, jaffres_spin_2010, fert_semiconductors_2007}. In a similar way we evaluate the spin heat resistance $r_s^Q$, which describes how effective a (non-equilibrium) spin heat accumulation $\Delta T_{s}$ is relaxed to zero in a non-magnetic material. Inelastic (electron-phonon and electron-electron) interactions are possible relaxation mechanisms. We shall include these later on, but we first discuss another mechanism, namely, elastic (or quasi-elastic) spin-flip scattering. The net effect is that it moves electrons above the Fermi energy from the hot spin reservoir to the cold spin reservoir, and simultaneously moves electrons below the Fermi energy in the opposite direction. This results in cooling of the hot spin reservoir and warming up of the colder spin reservoir, thus equalizing $T^{\uparrow}$ and $T^{\downarrow}$. Note that there is no net flow of spin angular momentum if the spin-flip scattering rate is not dependent on energy (i.e., unlike a spin accumulation, a non-zero $\Delta T_{s}$ does not give rise to spin relaxation, even though spin-flip scattering is the relevant scattering process. See the last subsection for an explicit evaluation).

The energy flow between the two spin reservoirs in the non-magnetic material is denoted by the spin heat current $J_s^{QV}$ per unit volume (in Wm$^{-3}$). We also introduce $\kappa_s^{V}$, the volume spin heat conductance (in Wm$^{-3}$K$^{-1}$) that connects the two spin reservoirs, such that
\begin{equation}
J_s^{QV} = \kappa_s^{V}\,\Delta T_{s} \label{eq26}
\end{equation}
The total energy flow is obtained by integrating $J_s^{QV}$ over the full spatial extent of the spin heat accumulation, noting that according to heat diffusion, the spin heat accumulation is expected to decay exponentially with vertical distance $z$ from the injection interface. That is, $\Delta T_{s} (z) = \Delta T_{s}\,exp(-z/\lambda_Q)$, where $\lambda_Q$ is the spin heat relaxation length. In a steady state, the integrated $J_s^{QV}$ has to be equal to the spin heat current $I_s^{Q}$ that is injected through the tunnel interface. Thus
\begin{eqnarray}
I_s^{Q} &=& \int_{0}^{\infty}J_s^{QV}\,dz = \kappa_s^{V}\int_{0}^{\infty}\Delta T_{s}(z)\,dz \nonumber\\
&=& \kappa_s^{V}\,\Delta T_{s}\,\lambda_Q \label{eq27}
\end{eqnarray}
Comparing this to the relation that defines $r_s^Q$ (Eq.~(\ref{eq12}) in the main text) we find that
\begin{equation}
r_s^Q = \frac{1}{2\,\kappa_s^{V}\,\lambda_Q} \label{eq28}
\end{equation}
Below it is shown that if the spin heat accumulation relaxes via elastic or quasi-elastic spin-flip scattering (only),  we have for $\kappa_s^{V}$
\begin{equation}
\kappa_s^{V} = \frac{L_0\,T_0}{2\,\rho\,(\lambda_s^{\text{el}})^2} = \frac{L_0\,T_0}{2\,r_s^{\text{el}}\,\lambda_s^{\text{el}}} \label{eq29}
\end{equation}
Because in that case also $\lambda_Q = \lambda_Q^{\text{el}} = \lambda_s^{\text{el}}$, we finally obtain
\begin{equation}
r_s^Q = \frac{r_s^{\text{el}}}{L_0\,T_0} \label{eq30}
\end{equation}
This is an important result, as it constitutes a type of spin-Wiedemann-Franz law, relating the electronic spin resistance $r_s$ to the spin heat resistance $r_s^Q$ via the Lorentz number and the temperature.

The final task is to prove Eq.~(\ref{eq29}) for $\kappa_s^{V}$. We assume that the spin-flip scattering is predominantly elastic. The spectral (energy-resolved) spin heat current due to spin-flip scattering, $J_s^{QV}(\varepsilon)$, in units of Wm$^{-3}$eV$^{-1}$, is then
\begin{equation}
J_s^{QV}(\varepsilon) = \frac{2\,\varepsilon\,N(\varepsilon)\left[f(\varepsilon,\mu^{\uparrow},T^{\uparrow})-f(\varepsilon,\mu^{\downarrow},T^{\downarrow})\right]}{\tau_{sf}^{\text{el}}(\varepsilon)} \label{eq31}
\end{equation}
where $\varepsilon$ is the energy with respect to the Fermi energy, $f(\varepsilon,\mu,T)$ is the Fermi-Dirac distribution function, $N(\varepsilon)$ is the density of states {\em per spin} in the non-magnetic material, and $\tau_{sf}^{\text{el}}$ is the spin-flip time. The factor of 2 appears because each spin flip reduces the energy difference between the two spin reservoirs by two units of $\varepsilon$. Note that $\tau_{sf}^{\text{el}}$ and the elastic spin-relaxation time $\tau_s^{\text{el}}$ are related by $\tau_{sf}^{\text{el}} = 2\,\tau_{s}^{\text{el}}$. If the density of states and the spin-flip time are not strongly dependent on energy, the spin heat current per unit volume is given by
\begin{eqnarray}
J_s^{QV} &=& \int_{-\infty}^{\infty}J_s^{QV}(\varepsilon)\,d\varepsilon  \label{eq32} \\
&=& \frac{N(\varepsilon_F)}{\tau_{s}^{\text{el}}} \int_{-\infty}^{\infty} \left[f(\varepsilon,\mu^{\uparrow},T^{\uparrow})-f(\varepsilon,\mu^{\downarrow},T^{\downarrow})\right] \varepsilon\,d\varepsilon \nonumber
\end{eqnarray}
The following relations are of use:
\begin{eqnarray}
f_i &=& f_0 + \frac{\partial f_0}{\partial \mu}(\mu_i - \mu_0) + \frac{\partial f_0}{\partial T}(T_i - T_0) \label{eq33}\\
\frac{\partial f_0}{\partial \mu} &=& -\frac{\partial f_0}{\partial \varepsilon} \label{eq34}\\
\frac{\partial f_0}{\partial T} &=& -\frac{\partial f_0}{\partial \varepsilon} \, \frac{\varepsilon}{T_0} \label{eq35}
\end{eqnarray}
where $f_0(\varepsilon,\mu_0,T_0)$ is the equilibrium distribution parameterized by $\mu_0$ and $T_0$. Using these relations we have
\begin{equation}
f(\varepsilon,\mu^{\uparrow},T^{\uparrow})-f(\varepsilon,\mu^{\downarrow},T^{\downarrow}) = -\frac{\partial f_0}{\partial \varepsilon}\Delta\mu -\frac{\partial f_0}{\partial \varepsilon} \frac{\varepsilon}{T_0} \Delta T_s \label{eq36}
\end{equation}
Inserting this into Eq.~(\ref{eq32}) gives
\begin{equation}
J_s^{QV} = \frac{N(\varepsilon_F)}{\tau_{s}^{\text{el}}} \int_{-\infty}^{\infty} \left[ -\frac{\partial f_0}{\partial \varepsilon}\,\varepsilon\,\Delta\mu -\frac{\partial f_0}{\partial \varepsilon} \, \frac{\varepsilon^2}{T_0} \,\Delta T_s \right] d\varepsilon
 \label{eq37}
\end{equation}
The integral can be evaluated using the Sommerfeld expansion (see Eq.~(3) of Hatami et al.\ Ref.~\onlinecite{hatami_thermoelectric_2009}):
\begin{eqnarray}
\int_{-\infty}^{\infty} \left(-\frac{\partial f_0}{\partial \varepsilon}\right) \varepsilon\,d\varepsilon &=& 0 \label{eq38} \\
\int_{-\infty}^{\infty} \left(-\frac{\partial f_0}{\partial \varepsilon}\right) \varepsilon^2\,d\varepsilon &=& \frac{\pi^2}{3}(kT_0)^2 = L_0\,T_0^2\,e^2
\label{eq39}
\end{eqnarray}
We then obtain
\begin{equation}
J_s^{QV} = \frac{N(\varepsilon_F)\,e^2}{\tau_{s}^{\text{el}}}\,L_0\,T_0\,\Delta T_s
 \label{eq40}
\end{equation}

Next we use the generalized Einstein relation, $\mu_{e}\,n^{tot}=D\,e\,(\partial n^{tot}/\partial E_F)$, where $\mu_{e}$ is the carrier mobility, $D$ the diffusion constant, $n^{tot}$ is the spin-integrated electron density, and $\partial n^{tot}/\partial E_F$ is equivalent to $2\,N(\varepsilon_F)$ (recall that the latter was defined per spin). Using the resistivity $\rho=1/(n^{tot}\,e\,\mu_{e})$, $\lambda_s^{\text{el}} = \sqrt{D\,\tau_s^{\text{el}}}$, and $r_s^{\text{el}} = \rho\,\lambda_s^{\text{el}}$, we obtain
\begin{equation}
J_s^{QV} = \frac{1}{2\,\rho\,(\lambda_s^{\text{el}})^2}\,L_0\,T_0\,\Delta T_s = \frac{L_0\,T_0}{2\,r_s^{\text{el}}\,\lambda_s^{\text{el}}}\,\Delta T_s
 \label{eq41}
\end{equation}
Comparing this to Eq.~(\ref{eq26}) gives
\begin{equation}
\kappa_s^{V} = \frac{L_0\,T_0}{2\,r_s^{\text{el}}\,\lambda_s^{\text{el}}} \label{eq42}
\end{equation}
just as it was already used in Eq.~(\ref{eq29}). Note that the spin heat current due to spin-flip scattering is proportional solely to $\Delta T_s$. There is no contribution from the non-zero $\Delta\mu$. This is a direct consequence of the assumption that the spin-flip time is not dependent on energy.

\subsection{Wiedemann-Franz law including inelastic relaxation}

In the previous subsection we derived a Wiedemann-Franz type of law for spin resistance, under the assumption that the relaxation of the spin heat accumulation occurs exclusively via (quasi-) elastic spin-flip scattering. However, spin heat relaxation occurs also via inelastic scattering processes.

One inelastic process is electron-phonon ($e$-$ph$) scattering. Phonons in the non-magnetic material cause an indirect energy flow from the hot spin reservoir to the cold spin reservoir, which can be understood as follows. In the presence of a non-zero $\Delta T_s$, the temperature difference $\Delta T_{e\text{-}ph}$ between electrons and phonons is spin dependent and given by $T^{\uparrow}-T^{ph}$ and $T^{\downarrow}-T^{ph}$, respectively. As a result, the heat transfer to the phonons is spin dependent, which tends to equalize $T^{\uparrow}$ and $T^{\downarrow}$. For the specific case where the phonon temperature is equal to the spin-averaged electron temperature $T_0$, we have $\Delta T_{e\text{-}ph}^{\uparrow} = +\Delta T_s\,/2$ and $\Delta T_{e\text{-}ph}^{\downarrow} = -\Delta T_s\,/2$.

Another process is inelastic electron-electron ($e$-$e$) scattering, which causes a direct energy flow between the two spin reservoirs in the non-magnetic material, which also tends to equalize $T^{\uparrow}$ and $T^{\downarrow}$. Note that in the absence of spin-orbit scattering \cite{heikkila_spin_2010} the $e$-$e$ interaction does not cause any additional relaxation of the spin accumulation $\Delta\mu$, so its sole effect is to decrease $\Delta T_{s}$.

Including relaxation via inelastic processes in the derivation of the spin-Wiedemann-Franz law has important effects. First, in Eq.~(\ref{eq26}) for the spin heat current $J_s^{QV}$ induced by a non-zero $\Delta T_s$, we must add the volume spin heat conductances $\kappa_s^{V,e\text{-}ph}$ and $\kappa_s^{V,e\text{-}e}$ due to  $e$-$ph$ and $e$-$e$ interactions, respectively. Therefore,
\begin{equation}
J_s^{QV} = \left( \kappa_s^{V} + \kappa_s^{V,e\text{-}ph} + \kappa_s^{V,e\text{-}e} \right) \Delta T_s
 \label{eq50}
\end{equation}
where we have kept $\kappa_s^{V}$ to denote the term due to elastic spin-flip scattering, given by Eq.~(\ref{eq42}).

The second effect is that we can no longer set $\lambda_s = \lambda_Q$, as previously done in Eq.~(\ref{eq28}) for the spin heat resistance. Let us discuss this in more detail. In the regime of diffusive (heat) transport, we expect that $\lambda_Q = \sqrt{D\,\tau_Q}$, with the corresponding diffusion constant being the same as that for charge transport. This is a valid assumption since \emph{electronic} heat transport is associated with electrons only. This is analogous to the case of electronic spin transport where the charge and spin diffusion coefficients coincide. On the other hand, the inelastic scattering processes contribute to the relaxation of the spin heat accumulation, and therefore affect $\tau_Q$. This spin heat relaxation time can be obtained by defining the inelastic contribution $\tau_Q^{\text{inel}}$ to it via
\begin{equation}
\kappa_s^{V,e\text{-}ph} + \kappa_s^{V,e\text{-}e} = \frac{N(\varepsilon_F)\,e^2\,L_0\,T_0}{\tau_Q^{\text{inel}}}
\end{equation}
such that, together with $\kappa_s^{V}=N(\varepsilon_F)\,e^2\,L_0\,T_0/\tau_s^{\text{el}}$ from Eq.~(\ref{eq40}), we have the total volume
spin heat conductance
\begin{equation}
\kappa_s^{V} + \kappa_s^{V,e\text{-}ph} + \kappa_s^{V,e\text{-}e} = \frac{N(\varepsilon_F)\,e^2\,L_0\,T_0}{\tau_Q}
\end{equation}
where
\begin{equation}
\frac{1}{\tau_Q} = \frac{1}{\tau_Q^{\text{el}}} + \frac{1}{\tau_Q^{\text{inel}}}
\end{equation}
Note that $\tau_Q^{\text{el}}=\tau_s^{\text{el}}$. Using Eq.~(\ref{eq40}) we can now write
\begin{equation}
\tau_Q = \tau_s^{\text{el}}\,\left(\frac{\kappa_s^{V}}{\kappa_s^{V} + \kappa_s^{V,e\text{-}ph} + \kappa_s^{V,e\text{-}e}}\right) \label{eq-tauq}
\end{equation}
For the case of $\kappa_s^{V} \gg \kappa_s^{V,e\text{-}ph} + \kappa_s^{V,e\text{-}e}$, this reduces to $\tau_s^{\text{el}}$.
To parameterize the contribution of inelastic scattering to the interspin energy exchange we define a parameter $\alpha$ as
\begin{equation}
\frac{1}{\alpha^2} = \frac{\tau_Q}{\tau_s} = \frac{\frac{1}{\tau_s^{\text{el}}} + \frac{1}{\tau_s^{\text{inel}}}}{\frac{1}{\tau_Q^{\text{el}}} + \frac{1}{\tau_Q^{\text{inel}}}} = \frac{\lambda_Q^2}{\lambda_s^2}
\end{equation}
The spin heat resistance including inelastic relaxation processes is obtained after Eq.~(\ref{eq28}) as
\begin{eqnarray}
r_s^Q &=& \frac{1}{2\left(\kappa_s^{V} + \kappa_s^{V,e\text{-}ph} + \kappa_s^{V,e\text{-}e}\right)\lambda_Q}\\
&=& \frac{\tau_Q}{2\,\tau_s^{\text{el}}\,\kappa_s^{V}\,\lambda_Q} \nonumber \\
&=& \frac{r_s^{\text{el}}}{L_0\,T_0}\,\frac{\tau_Q}{\tau_s^{\text{el}}}\,\frac{\lambda_s^{\text{el}}}{\lambda_Q} \nonumber \\
&=& \frac{r_s^{\text{el}}}{L_0\,T_0}\,\frac{\lambda_Q}{\lambda_s^{\text{el}}} \nonumber
\end{eqnarray}
where we have first inserted Eq.~(\ref{eq-tauq}), then used Eq.~(\ref{eq42}) to eliminate $\kappa_s^{V}$ and then used that
$\tau_Q / \tau_s^{\text{el}} = (\lambda_Q / \lambda_s^{\text{el}})^2$. Finally, expressing the spin resistance $r_s^{\text{el}}=\rho\,\lambda_s^{\text{el}}$
due to elastic processes only in terms of the total spin resistance $r_s=\rho\,\lambda_s$ including inelastic processes,
we finally obtain
\begin{equation}
r_s^Q = \frac{r_s}{L_0\,T_0}\times\frac{1}{\alpha}
\end{equation}
which is the spin-Wiedemann-Franz law as described in Eq.~(\ref{eq22}) of the main text.

\subsection{Spin relaxation with finite spin heat accumulation}

For the sake of completeness, we evaluate the spin current due to elastic spin-flip scattering in the presence of a spin heat accumulation using the same approach as in the first subsection. It is demonstrated that a non-zero $\Delta T_s$ does not induce nor change the spin accumulation, provided that the spin-flip time (and the density of states around the Fermi level) are not dependent on energy in the relevant range of a few $kT$. The spectral spin current density per unit volume (in Am$^{-3}$) is
\begin{equation}
J_s^{V}(\varepsilon) = \frac{2\,N(\varepsilon)\left[f(\varepsilon,\mu^{\uparrow},T^{\uparrow})-f(\varepsilon,\mu^{\downarrow},T^{\downarrow})\right]\,e}{\tau_{sf}^{\text{el}}(\varepsilon)} \label{eq43}
\end{equation}
If $\tau_{sf}^{\text{el}}$ and $N$ do not vary much around the Fermi energy, the integrated spin current in the presence of a non-zero $\Delta\mu$ and a non-zero $\Delta T_s$ is
\begin{equation}
J_s^{V} =
\frac{N(\varepsilon_F)\,e}{\tau_{s}^{\text{el}}} \int_{-\infty}^{\infty} \left[f(\varepsilon,\mu^{\uparrow},T^{\uparrow})-f(\varepsilon,\mu^{\downarrow},T^{\downarrow})\right] d\varepsilon \label{eq44}
\end{equation}
Using Eq.~(\ref{eq36}) as in the first subsection, this is rewritten as
\begin{equation}
J_s^{V} = \frac{N(\varepsilon_F)\,e}{\tau_{s}^{\text{el}}} \int_{-\infty}^{\infty} \left[-\frac{\partial f_0}{\partial \varepsilon}\,\Delta\mu -\frac{\partial f_0}{\partial \varepsilon}\, \frac{\varepsilon}{T_0}\,\Delta T_s\right] d\varepsilon
\label{eq45}
\end{equation}
The second part of the integral, which contains $\Delta T_s$ vanishes (see Eq.~(3) of Hatami et al.\ Ref.~\onlinecite{hatami_thermoelectric_2009}). Thus, only a non-zero $\Delta\mu$ produces a net spin current and spin relaxation. Now, using
\begin{equation}
\int_{-\infty}^{\infty} \left(-\frac{\partial f_0}{\partial \varepsilon}\right)\,d\varepsilon = 1
\label{eq46}
\end{equation}
and
\begin{equation}
\frac{N(\varepsilon_F)\,e}{\tau_{s}^{\text{el}}} = \frac{1}{2\,e\,\rho\,(\lambda_s^{\text{el}})^2}
\label{eq47}
\end{equation}
we obtain
\begin{equation}
J_s^{V} = \frac{1}{2\,e\,\rho\,(\lambda_s^{\text{el}})^2}\,\Delta\mu
\label{eq48}
\end{equation}
Assuming that $\Delta\mu (z) = \Delta\mu\,exp(-z/\lambda_s)$, and using that in a steady state, the spatially integrated $J_s^{V}$ has to be equal to the spin current $I_s$ that is
injected through the tunnel interface, we obtain
\begin{eqnarray}
I_s &=& \int_{0}^{\infty}J_s^{V}\,dz = \frac{1}{2\,e\,\rho\,(\lambda_s^{\text{el}})^2}\int_{0}^{\infty}\Delta\mu(z)\,dz \nonumber\\
&=& \frac{1}{2\,e\,\rho\,\lambda_s^{\text{el}}}\Delta\mu = \frac{1}{2\,e\,r_s^{\text{el}}}\Delta\mu \label{eq49}
\end{eqnarray}
where $r_s^{\text{el}}=\rho\,\lambda_s^{\text{el}}$ is the spin resistance (in $\Omega$m$^2$), consistent with the definition of $r_s$ in Eq.~(\ref{eq11}) of the main text.

\section{Hanle effect and Bloch equations}
In this section we describe the dynamics of the spin accumulation and the spin heat accumulation due to the Hanle effect.
A spin accumulation in a paramagnetic material implies that there is a net non-equilibrium magnetization or particle spin density $N_s$. The particle spin density is linearly proportional to the spin accumulation $\Delta\mu$ if the latter is sufficiently small:
\begin{eqnarray}
N_s &=& (N^{\uparrow} - N^{\downarrow})\,\frac{\hbar}{2} \nonumber \\
&=& \frac{\hbar}{2}\,\int_{-\infty}^{\infty} N(\varepsilon) \left[f(\varepsilon,\mu^{\uparrow},T_n^{\uparrow})-f(\varepsilon,\mu^{\downarrow},T_n^{\downarrow})\right] d\varepsilon \nonumber \\
&=& N(\varepsilon_F)\,\frac{\hbar}{2} \int_{-\infty}^{\infty} \left[\left(-\frac{\partial f_0}{\partial \varepsilon}\right)\,\Delta\mu + \left(-\frac{\partial f_0}{\partial \varepsilon}\right) \frac{\varepsilon}{T} \,\Delta T_s \right]\,d\varepsilon \nonumber \\
&=& N(\varepsilon_F)\,\frac{\hbar}{2}\,\Delta\mu
\end{eqnarray}
Similarly, if we denote the energy density by $E$, then the energy spin density $E_{s} = E^{\uparrow} - E^{\downarrow}$ is proportional to the spin heat accumulation $\Delta T_s$:
\begin{eqnarray}
E_{s} &=& E^{\uparrow} - E^{\downarrow} \nonumber \\
&=&  \int_{-\infty}^{\infty} N(\varepsilon) \left[f(\varepsilon,\mu^{\uparrow},T_n^{\uparrow})-f(\varepsilon,\mu^{\downarrow},T_n^{\downarrow})\right] \varepsilon\,d\varepsilon \nonumber \\
&=& N(\varepsilon_F) \int_{-\infty}^{\infty} \left[\left(-\frac{\partial f_0}{\partial \varepsilon}\right)\,\varepsilon\,\Delta\mu + \left(-\frac{\partial f_0}{\partial \varepsilon}\right) \frac{\varepsilon^2}{T} \,\Delta T_s \right]\,d\varepsilon \nonumber \\
&=& N(\varepsilon_F)\,e^2\,L_0\,T_0 \,\Delta T_s
\end{eqnarray}
When a magnetic field is applied transverse to the initial spin direction (Hanle geometry), the electron magnetic moments precess at the Larmor frequency. In the next paragraph we first describe the resulting dynamics of the particle spin density using the Bloch equations, and then apply a similar analysis to the energy spin density to describe the dynamics of the spin heat accumulation.

The Bloch equations that describe the dynamics (precession and relaxation) of the particle spin density are:
\begin{eqnarray}
\frac{\partial N_{s,x}}{\partial t} &=& \gamma (N_s \times B)_x - \frac{N_{s,x}}{\tau_s} \\
\frac{\partial N_{s,y}}{\partial t} &=& \gamma (N_s \times B)_y - \frac{N_{s,y}}{\tau_s} \\
\frac{\partial N_{s,z}}{\partial t} &=& \gamma (N_s \times B)_z - \frac{N_{s,z}}{\tau_s}
\end{eqnarray}
with $\gamma = g\,\mu_B/\hbar$ the gyromagnetic ratio. For $B$ applied along the z-axis, 
and the boundary conditions $N_{s,x}=N_0$ and $N_{s,y}=N_{s,z}=0$ at $t=0$, the solutions are:
\begin{eqnarray}
N_{s,x} &=& N_{0}\,cos(\omega_L\,t)\,e^{-t/\tau_s} \label{eq:bloch1} \\
N_{s,y} &=& N_{0}\,sin(\omega_L\,t)\,e^{-t/\tau_s} \label{eq:bloch2} \\
N_{s,z} &=& 0 \label{eq:bloch3} 
\end{eqnarray}
Thus, the precession of the magnetic moments in a transverse magnetic field causes the projection of the particle spin density onto the x and y axis to oscillate at the Larmor frequency, whereas spin relaxation causes an exponential decay. Since the particle spin density has a magnitude and a direction that varies in time, and the particle spin density is proportional to the spin accumulation, the latter also has a magnitude and direction that varies in time. Integrating over time gives a Lorentzian for the x-component of the particle spin density,
and similarly for the x-component of the spin accumulation:
\begin{equation}
N_{s,x} \propto (\Delta\mu)_{x} = (\Delta\mu)_0\,\frac{1}{1+(\omega_L\,\tau_s)^2}
\end{equation}

The energy spin density $E_{s} = E^{\uparrow} - E^{\downarrow}$ describes the difference in the energy of electrons with spin pointing parallel ($\uparrow$) and antiparallel ($\downarrow$) to a given quantization axis. The precession of the electron magnetic moments in a transverse magnetic field causes the quantization axis to be time dependent, and hence the energy spin density also precesses at the Larmor frequency. To describe this, we introduce a similar set of Bloch equations, but now for the energy spin density, and taking into account that a different relaxation time $\tau_Q$ should be used:
\begin{eqnarray}
\frac{\partial E_{s,x}}{\partial t} &=& \gamma (E_s \times B)_x - \frac{E_{s,x}}{\tau_Q} \\
\frac{\partial E_{s,y}}{\partial t} &=& \gamma (E_s \times B)_y - \frac{E_{s,y}}{\tau_Q} \\
\frac{\partial E_{s,z}}{\partial t} &=& \gamma (E_s \times B)_z - \frac{E_{s,z}}{\tau_Q}
\end{eqnarray}
Please note that the energy spin density can be defined even if the particle spin density is zero. For $B$ applied along the z-axis, and with similar boundary conditions ($E_{s,x}=E_0$ and $E_{s,y}=E_{s,z}=0$ at $t=0$) the solutions are thus similar to Eqs.~(\ref{eq:bloch1}), (\ref{eq:bloch2}) and (\ref{eq:bloch3}), namely:
\begin{eqnarray}
E_{s,x} &=& E_{0}\,cos(\omega_L\,t)\,e^{-t/\tau_Q}\\
E_{s,y} &=& E_{0}\,sin(\omega_L\,t)\,e^{-t/\tau_Q}\\
E_{s,z} &=& 0
\end{eqnarray}
Since the energy spin density has a magnitude and a direction that varies in time, and it is proportional to the spin heat accumulation, the latter also has a magnitude and direction that varies in time, effectively making the spin heat accumulation a (time-dependent) vector. Integrating over time gives a Lorentzian for the x-component:
\begin{equation}
(\Delta T_{s})_{x} = (\Delta T_{s})_0\,\frac{1}{1+(\omega_L\,\tau_Q)^2}
\end{equation}

In the main text it was shown that the voltage across the tunnel contact has two independent contributions from, respectively, the spin accumulation and the spin heat accumulation. This applies to the static case as well as under dynamic conditions. The Hanle voltage signal is thus a superposition of two voltage signals:
\begin{equation}
\Delta V_{Hanle} = \left(\frac{P_G}{2\,e}\right)\Delta\mu + \left(\frac{P_L\,S}{2}\right)\Delta T_{s}
\end{equation}
The Hanle line shape is thus also a superposition of two Lorentzian curves with different amplitude and line width, determined by $\Delta\mu$ and $\Delta T_{s}$ and by $\tau_s$ and $\tau_Q$:
\begin{eqnarray}
\Delta V_{Hanle}(B_z) &=& \left(\frac{P_G}{2\,e}\right)\,\frac{(\Delta\mu)_0}{1+(\omega_L\,\tau_s)^2} \nonumber \\
&&+ \left(\frac{P_L\,S}{2}\right)\,\frac{(\Delta T_{s})_0}{1+(\omega_L\,\tau_Q)^2}
\end{eqnarray}

\end{document}